\documentclass[%
 reprint,
superscriptaddress,
 amsmath,amssymb,
 aps,
]{revtex4-1}

\usepackage[colorlinks,plainpages=false,linkcolor=blue,urlcolor=blue,citecolor=blue,pdfpagemode=UseNone,pdfstartview=FitBH]{hyperref}

\usepackage{graphicx}
\usepackage{dcolumn}
\usepackage{bm}

\begin{document}

Disclaimer: This manuscript has been authored by UT-Battelle, LLC under Contract No. DE-AC05-00OR22725 with the U.S. Department of Energy. The United States Government retains and the publisher, by accepting the article for publication, acknowledges that the United States Government retains a non-exclusive, paid-up, irrevocable, world-wide license to publish or reproduce the published form of this manuscript, or allow others to do so, for United States Government purposes. The Department of Energy will provide public access to these results of federally sponsored research in accordance with the DOE Public Access Plan (http://energy.gov/downloads/doe-public-access-plan).
\clearpage
\setcounter{page}{0}

\preprint{APS/123-QED}

\title{Inelastic neutron scattering study of the anisotropic $S = 1$ spin chain [Ni(HF$_2$)(3-Clpyridine)$_4$]BF$_4$}

\author{Daniel M. Pajerowski}
 \email{pajerowskidm@ornl.gov}
 \affiliation{Neutron Scattering Division, Oak Ridge National Laboratory, Oak Ridge, TN 37831, USA}
\author{Jamie L. Manson}%
 \affiliation{Department of Chemistry and Biochemistry, Eastern Washington University, Cheney, WA 99004, USA}%
\author{Jacek Herbrych}
 \affiliation{Department of Theoretical Physics, Wrocław University of Science and Technology, 50-370 Wrocław, Poland}
 \affiliation{Department of Physics and Astronomy, University of Tennessee, Knoxville, TN 37996, USA}
 \affiliation{Materials Science and Technology Division, Oak Ridge National Laboratory, Oak Ridge, TN 37831, USA}
\author{Jesper Bendix}
 \affiliation{Department of Chemistry, University of Copenhagen, Universitetsparken 5, DK-2100 Copenhagen, Denmark}
\author{Andrey P. Podlesnyak}
 \affiliation{Neutron Scattering Division, Oak Ridge National Laboratory, Oak Ridge, TN 37831, USA}
\author{John M. Cain}
 \affiliation{Department of Chemistry, University of Florida, Gainesville, FL 32611-8440, USA}
\author{Mark W. Meisel}
 \affiliation{Department of Physics and the National High Magnetic Field Laboratory, University of Florida, Gainesville, FL 32611-8440, USA}

\date{\today}

\begin{abstract}
[Ni(HF$_2$)(3-Clpyridine)$_4$]BF$_4$ (NBCT) is a one-dimensional, $S = 1$ spin chain material that shows no magnetic neutron Bragg peaks down temperatures of 0.1~K.  Previous work identified NBCT to be in the Haldane phase and near a quantum phase transition as a function of $D/J$ to the large-$D$ quantum paramagnet phase (QPM), where $D$ is the axial single-ion anisotropy and $J$ is the intrachain superexchange.  Herein, inelastic neutron scattering results are presented on partially deuterated, $^{11}$B enriched NBCT polycrystalline samples in zero magnetic field and down to temperatures of 0.3~K.  Comparison to density matrix renormalization group calculations yields $D/J = 1.51$ and a significant rhombic single-ion anisotropy $E$ ($E/D \approx 0.03$, $E/J \approx 0.05$).  These $D$, $J$, and $E$ values place NBCT in the large-$D$ QPM phase but precipitously near a quantum phase transition to a long-range ordered phase.
\end{abstract}

\maketitle


\section{\label{introduction}Introduction\protect\\}
Spin chains have played a foundational role in understanding many-body physics in the quantum regime, dating to nearly a century ago and continue to yield interesting physics.  Lower spin values naturally possess more quantum features, and the $S = 1/2$ and $S = 1$ isotropic chains have distinct ground-states \cite{Affleck1989}.  The $S = 1/2$ class of materials is unique in that it may be considered analytically via Bethe’s approach, giving a gapless ground-state \cite{Bethe1931}. Conversely, the isotropic $S = 1$ antiferromagnetic (AFM) spin chains do not have analytical solutions and have a non-degenerate gapped ground-state that is called the Haldane phase \cite{Haldane1983prl,Haldane1983,Haldane2017}. These spin chains are also notable as prototypes for considering topologically-ordered physics \cite{Wen2017}, having a hidden nonlocal order parameter \cite{Nijs1989}.

The present work focuses on the effects of single-ion anisotropy on the ground-state and excitations of $S = 1$ chains that may be characterized by the spin Hamiltonian
\begin{eqnarray}
H = J\sum_{i} \mathbf{S}_i \cdot \mathbf{S}_{i+1} + J'\sum_{<i,j>} \mathbf{S}_i \cdot \mathbf{S}_{j} \nonumber\\
+ D\sum_{i} (S_i^z)^2 + E\sum_{i} [(S_i^x) - (S_i^y)]^2,
\label{eq:1}
\end{eqnarray}
where $\mathbf{S}_i=(S^x_i,S^y_i,S^z_i)$, $J > 0$ is the AFM intrachain superexchange energy, $J'$ is the interchain superexchange energy and the $<i,j>$ summation is between neighboring chains, $D$ is the single-ion axial anisotropy, and $E$ is the single-ion rhombic anisotropy.  Theoretical consideration and numerical studies of eq. \eqref{eq:1} gives rise to a complex phase diagram in $J$, $J'$, $D$, and $E$ with an array of quantum critical boundaries and quantum multi-critical points.\cite{Hu2011,Chen2003,Tzeng2017,Albuquerque2009}  An additional term of exchange anisotropy has also been considered in the so-called $XXZ$ chains \cite{Hu2011,Chen2003}. Only more recently has the rhombic anisotropy been considered, and like $D$ it destabilizes the Haldane phase \cite{Tzeng2017}.  For $J'= 0$ and $E = 0$ without exchange anisotropy, the Haldane phase is bounded by the critical easy-plane anisotropy $D_C/J = 0.96845$ \cite{Hu2011} and the critical easy-axis anisotropy $D_C/J = -0.32$ \cite{Chen2003,Albuquerque2009}. 

From the perspective of the dynamical (energy-resolved) correlation functions, spin chains described by eq. \eqref{eq:1} have a singlet ground state with propagating magnetic modes as excited states. For the isotropic case ($D=E=0$) there is one mode of spin-spin correlations $\langle S^\alpha S^\alpha \rangle$ (with $\alpha=x,y,z$) that has a gap at the AFM zone center (often called the $\pi$-point) of $\Delta = 0.41 J$ \cite{White1993}. Introducing $D$ splits the propagating modes into longitudinal $\langle S^zS^z \rangle$ and transverse $\langle S^xS^x \rangle=\langle S^yS^y \rangle$ components. Finally, finite $E$ splits the transverse mode, $\langle S^xS^x \rangle\ne\langle S^yS^y \rangle$.

Experimentally, compounds having various values of $J$, $J'$, $D$, $E$ have been reported.  A summary of experimentally realized $S = 1$ spin chains for different $J$, $J'$, and $D$ may be found in Table 1 of reference \cite{Wierschem2014}, although $E$ is not included there.  It stands out that there is a large number of compounds that are well within the Haldane phase with $D/J < 0.25$ and compounds with $D/J > 4$ that are in the large-$D$ quantum paramagnet (QPM) phase.  It is exceptional that [Ni(HF$_2$)(3-Clpyridine)$_4$]BF$_4$ (NBCT) is projected to have $D/J \approx D_C/J \approx 1$.

\begin{figure}
\includegraphics{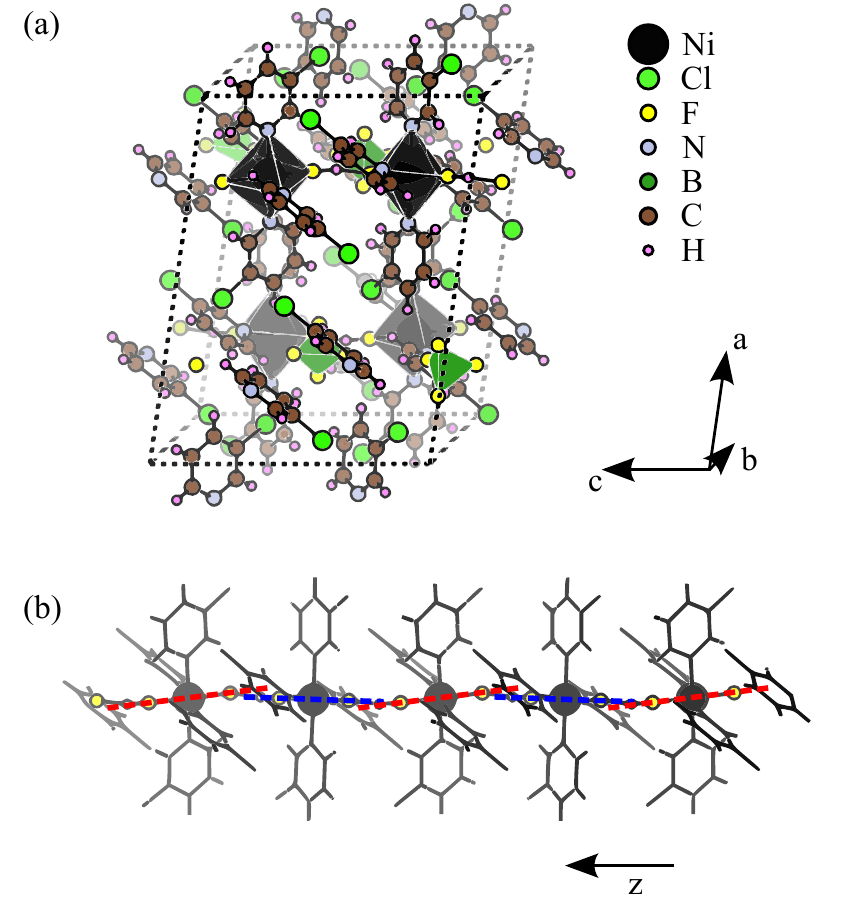}
\caption{\label{fig:crystal_structure} NBCT crystal structure and chain motif. (a) The NBCT unit cell, NiF$_2$N$_4$ (black) and BF$_4$ polyhedra (green) are shown.  (b) A chain portion along the crystallographic c-axis and the z-axis of eq. \eqref{eq:1}, with atoms not linking the chain suppressed.  The dashed lines illustrate the relative canting between the two NiF$_2$N$_4$ moeities along the z-(c-)axis.}
\end{figure}

The NBCT system is a coordination polymer with $S$ = 1 Ni$^{2+}$ magnetic ions octahedrally coordinated as NiF$_2$N$_4$, where the nickel chains are separated by large chloropyrizine ligands and the intra-chain nickel bonds are via Ni-F-H-F-Ni linkages, Fig.~\ref{fig:crystal_structure} (a) \cite{Manson2012}. The physical chain separation of $>10$ \AA\ without any apparent electron hopping pathways makes a strong case for $J' \approx 0$ in NBCT.  There is a $34^{\circ}$ canting angle between the unique axes of the nickel along the chain, Fig.~\ref{fig:crystal_structure} (b).  The unit cell parameter along the chain, c-axis is 12.291 \AA\ and there are two nickel ions within the unit cell along this direction such that the $\pi$-point is at a momentum of 0.511 \AA$^{-1}$.  The initial identification of $D/J$ = 0.88 for NBCT was derived by fitting the high-temperature susceptibility, assuming $D$ = 0, to obtain $J$ = 4.86~K (0.42~meV) and $g$ = 2.10, and  UV-Vis data to extract $D$ = 4.3~K (0.37~meV).  No $E$ value has been reported.  The proximity of NBCT to a phase boundary was also inferred from isothermal magnetization studies that used randomly arranged microcrystals cooled down to 50 mK, and an upper limit for a possible critical magnetic field (which could indicate existence of the gap) was given as $H_C \lesssim 35 \pm 10$ mT  \cite{Xia2018}.

Here, to further characterize the NBCT material, we have used inelastic neutron scattering (INS) experiments and density matrix renormalization group (DMRG) \cite{Schollw} calculations. The INS technique can directly probe time and space spin correlations in materials and has therefore been extensively used to investigate spin chain materials \cite{Broholm2001}. In the following, we use INS from isotopically enriched powder samples to quantify the collective magnetic modes in NBCT at $T$ = 0.3~K. The observed spectra are then compared with $T=0$ DMRG calculations of the dynamical spin structure factor $S(q,\omega)$ of eq. \eqref{eq:1} with $J'=0$ to extract $D$, $J$, and $E$. The important findings are that $D/J$ = 1.51 and $E/J$ $\approx$ 0.05 for our isotopically doped sample of NBCT. These results are in striking contrast to the $D/J$ = 0.88 value reported for NBCT without isotope doping using bulk probes.  The concluding section frames the present results in the context of the existing literature and provides suggestions for future studies of this unique system. Additional technical details are given in the Appendices.

\section{\label{results_and_discussion}Results and Discussion\protect}
\begin{figure*}
\includegraphics{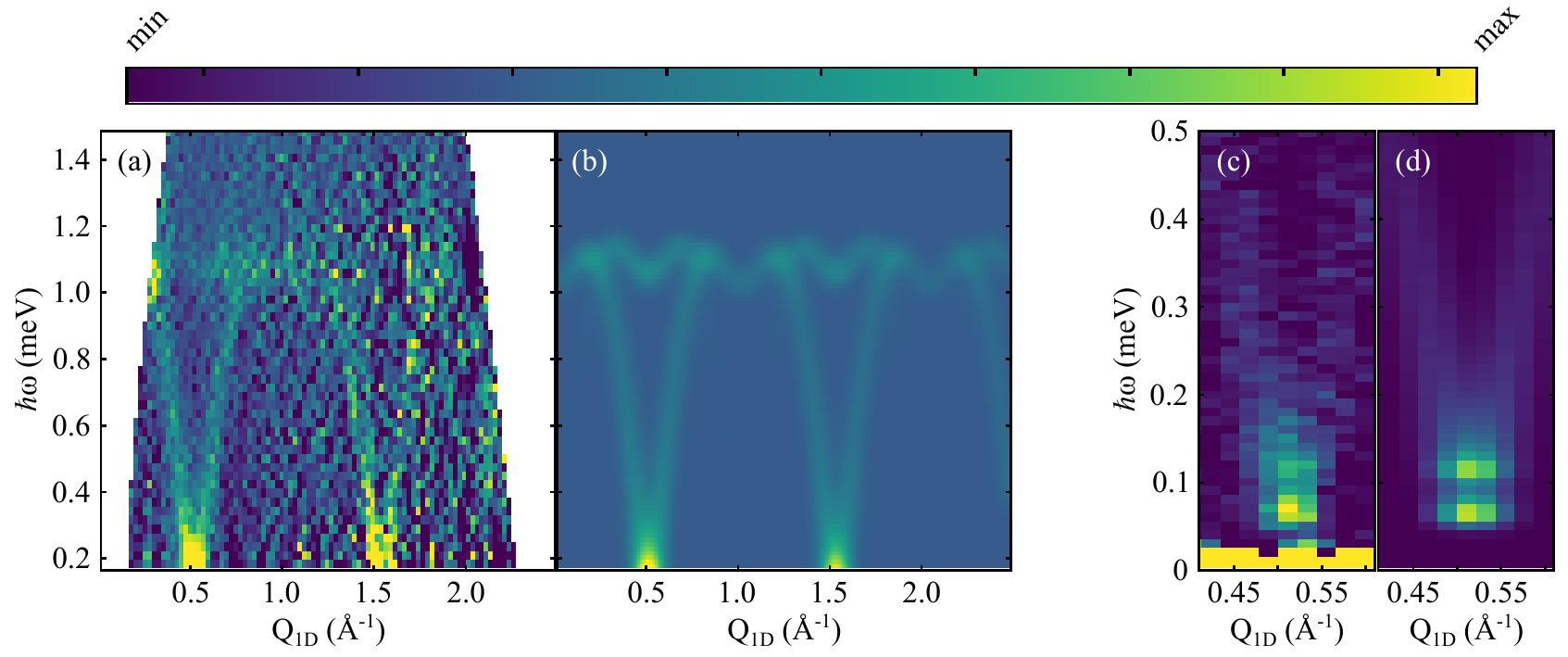}
\caption{\label{fig:neutron_maps}Intensity maps of reverse-power-averaged NBCT experimental data and DMRG model.  (a)  E$_i$ = 3.32~meV, $T$ = 0.3~K data minus $T$ = 13~K data.  White regions are where the scattering condition is not satisfied by the spectrometer.  (b) E$_i$ = 3.32~meV model that is calculated at $T$ = 0, but includes a $T$ = 0.3~K Bose factor correction.  For (a) and (b) the intensity scale has min = $–2\times10^{-4}$ and max = $5\times10^{-4}$ with scaled units.  (c) E$_i$ = 1.00~meV, $T$ = 0.3~K data.  (d) E$_i$ = 1.00~meV that is calculated at $T$ = 0, but includes a $T$ = 0.3~K Bose factor correction.  For (b) and (d) the intensity scale has min = 0 and max = $2\times10^{-4}$ with scaled units.}
\end{figure*}

Neutron spectra were collected at $T$ = 0.07~K and showed no additional magnetic Bragg scattering (e.g. no magnetic ordering down to that temperature), but the background multiple scattering from that setup proved problematic for measuring inelastic features.  Additional INS measurements were performed with a different cryostat at $T$ = 0.3~K and 13~K for incident energies (E$_i$’s) of 1.00~meV, 1.55~meV, 3.32~meV, and 7.00~meV.  No magnetic signal was observed above energy transfers of $\hbar \omega \approx$ 1.2~meV.  The one-dimensional scattering function for NBCT was extracted from the powder data using the reported method \cite{Tomiyasu2009}. For the E$_i$ = 3.32~meV data, the dispersionless $T$ = 13~K data were subtracted from the $T$ = 0.3~K data to remove extrinsic multiple scattering signatures, while the E$_i$ = 1.00~meV data have no such subtraction.  A lattice-periodic dispersive mode is visible in the E$_i$ = 3.32~meV data, Fig. \ref{fig:neutron_maps} (a).  Two gaps at $q_{1D}$ = $\pi$ ($Q_{1D} \approx 0.51 $Å$^{-1}$) visible in the E$_i$ = 1.00~meV data, Fig. \ref{fig:neutron_maps} (c), with values of $\Delta_1$ = 0.057~meV and $\Delta_2$ = 0.111~meV from the fits described below.  The momentum transfer along the chain is $Q_{1D}$ and the unitless momentum $q_{1D}$ varies from 0 to 2$\pi$ in the Brillouin zone.

To extract Hamiltonian parameters, the experimental data were compared to DMRG calculations.  For the E$_i$ = 3.32~meV data, the optimization region is between $\hbar\omega$ = [0.8, 1.3]~meV and $Q_{1D} <$ 1 Å$^{-1}$.  For the E$_i$ = 1.00~meV data, the optimization region is between $\hbar\omega$ = [0.045, 0.400]~meV and $Q_{1D}$ = [0.44, 0.60] Å$^{-1}$.  Initial conditions for the fitting were $D/J$ = [0, 0.5, 1.0, 1.5, 2.0] with $J$ values initialized to have the model zone boundary peak intensity at $\hbar\omega$ = 1.1~meV.  These fits have three intrinsic parameters $D$, $J$ and $\Delta_E$, where $\Delta_E$ captures the effect of the rhombic $E$-term by modifying that gap of the $\langle S^{x/y}S^{x/y}\rangle$ mode $\Delta_{x/y} \rightarrow \Delta_{x/y} \pm \Delta_E$/2 with $\Delta_{x/y} = (\Delta_1 + \Delta_2)/2 = 0.084$~meV.  There is one extrinsic parameter that scales the overall intensity.  Relative intensity between the E$_i$ = 3.32~meV and E$_i$ = 1.00~meV was taken from the known flux difference of those energies, the momentum resolution was taken from Bragg peaks, and the energy resolution used semi-empirical model that has been developed for the spectrometer.  The lowest residuals were found with the $D/J = 1.51$ parameters in Table I, which yields the spectra in Figures \ref{fig:neutron_maps} (b) and (d).

\begin{figure}
\includegraphics{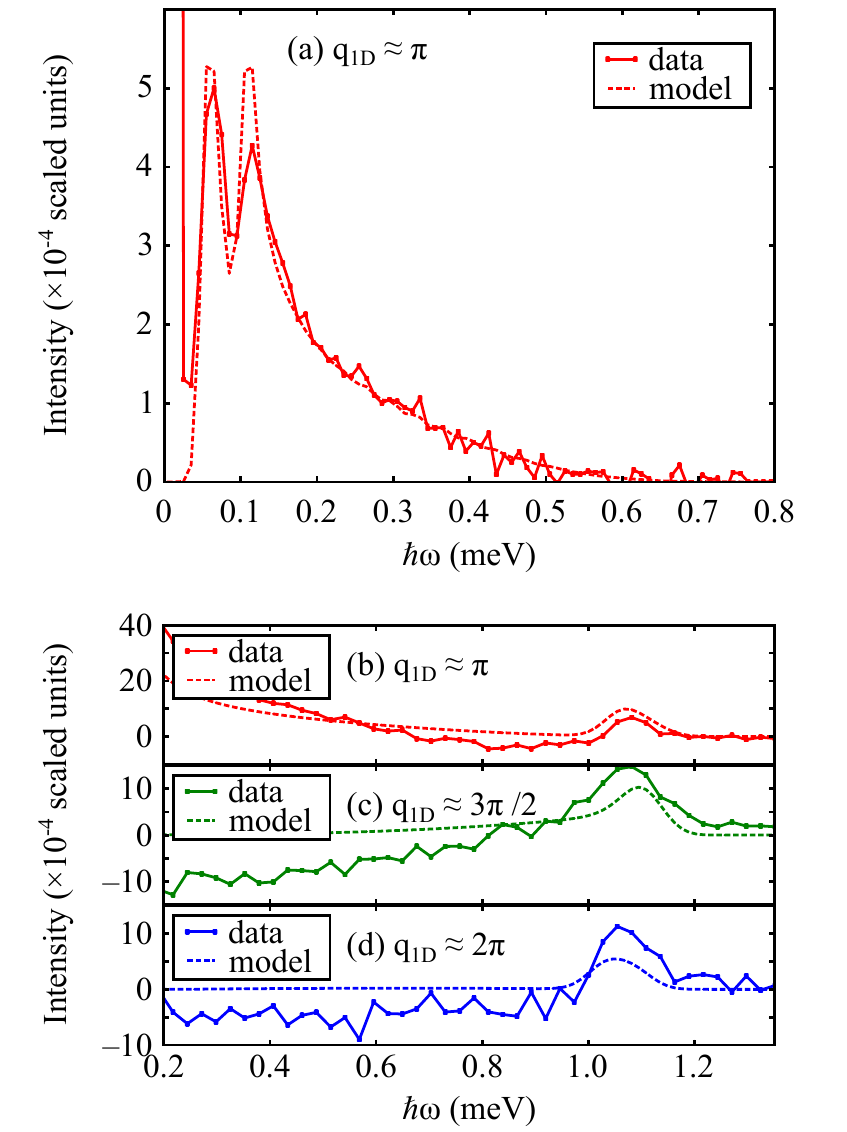}
\caption{\label{fig:neutron_line} Intensity versus energy transfer for reverse-power-averaged NBCT experimental data and DMRG model.  (a) Data and model of E$_i$ = 1.00~meV, $T$ = 0.3~K are shown for $Q_{1D}$ = $[0.44, 0.57]$ Å$^{-1}$.  Data and model of E$_i$ = 3.32~meV, $T$ = 0.3~K are shown for (b) $Q_{1D}$ = $[0.41, 0.65]$ Å$^{-1}$ centered on the $\pi$-point, (c) $Q_{1D}$ = $[0.66, 0.90]$ Å$^{-1}$ centered on the $3\pi/2$-point, and (d) $Q_{1D}$ = $[0.91, 1.15]$ Å$^{-1}$ centered on the 2$\pi$-point.  The negative intensities in the E$_{i}$ = 3.32~meV are most likely due to over subtraction of the $T$ = 13~K data that has quasi-elastic magnetic scattering.}
\end{figure}

A more quantitative visualization of the best-fit model compared to the data is possible by integrating over some momentum regions.  The splitting of the gap at $q_{1D}$ = $\pi$ and quantitative gap-energy fitting of the model to the data is shown in Fig. \ref{fig:neutron_line}.  Some higher energy lines further illustrate the good quantitative agreement of the $D/J$ = 1.51 model with the INS data, Fig. \ref{fig:neutron_line} (b-d).  This higher energy spectral weight is essential to the model, as without it the gap energies have no context, and there is also information regarding the dispersivity of the $\langle S^zS^z \rangle$ mode.  It is unclear if the incorrect model intensity in Fig. \ref{fig:neutron_line} (a) shows an intrinsic effect or is merely due to a difference between the recorded thermometer temperature and the actual sample temperature (a temperature of $T \approx$ 0.6~K would increase the model intensity of the lowest mode and reproduce the experimentally observed ratio of the two modes).  Indeed, it may be that the difficulty in cooling is due to a large entropy contribution of the ground state.

When starting INS model optimization with the $D/J$ = 0.5 initial condition, a local minimum exists having residuals $>$10\% larger than the best-fit that is a similar distance in $D/J$ from the critical $D_C/J$ value, and with parameters of $J$ = 0.41~meV, $D$ = 0.18~meV, $D/J$ = 0.44, and $\Delta_E$ = 0.05~meV.  While fitting the gap well, the $D/J$ = 0.44 fit is qualitatively different than the data as it has an $\langle S^zS^z \rangle$ mode with significant intensity at ($q_{1D}$, $\hbar\omega$) = ($\pi$, 0.44~meV) that would be greater than five times the observation along with missing the intensity at the top of the band at the AFM zone boundary.

\begin{table}[b]
\caption{\label{tab:table1}%
Best-fit parameters of NBCT.  Uncertainties are from fitting to least-squares.
}
\begin{ruledtabular}
\begin{tabular}{lcccc}
\textrm{}&
\textrm{J (meV)}&
\textrm{D (meV)}&
\textrm{D/J}&
\textrm{$\Delta_E$ (meV)}\\
\colrule
NBCT & 0.35 $\pm$ 0.01 & 0.53 $\pm$ 0.01 & 1.51 $\pm$ 0.01 & 0.05 $\pm$ 0.01\\
\end{tabular}
\end{ruledtabular}
\end{table}

In addition to the one-dimensional collective modes, a local mode was observed at low-temperatures, Appendix B.  This local mode is not visible in the data converted from $|Q|$ to $Q_{1D}$ because it does not have a one-dimensional momentum dependence of the intensity.  In the unconverted data, there is a momentum independent (aside from the Ni$^{2+}$ magnetic form factor) peak at 0.53~meV that is precisely the best-fit value of $D$.  We assign this feature to be single-ion $D$ excitations from Ni$^{2+}$ spins that are in environments that have chain lengths less than the correlation length.

Our neutron data give different parameters than were derived from bulk measurements, but have values that are close to being within the bulk parameter experimental uncertainties.  The $D/J$ = 0.44 solution has $D$ = 0.18~meV, which is low for NiF$_2$N$_4$ \cite{Manson2020}.  Then with $D/J$ = 1.51, NBCT is in the large-$D$ QPM phase in  $(D,J)$-space.

\begin{figure}
\includegraphics{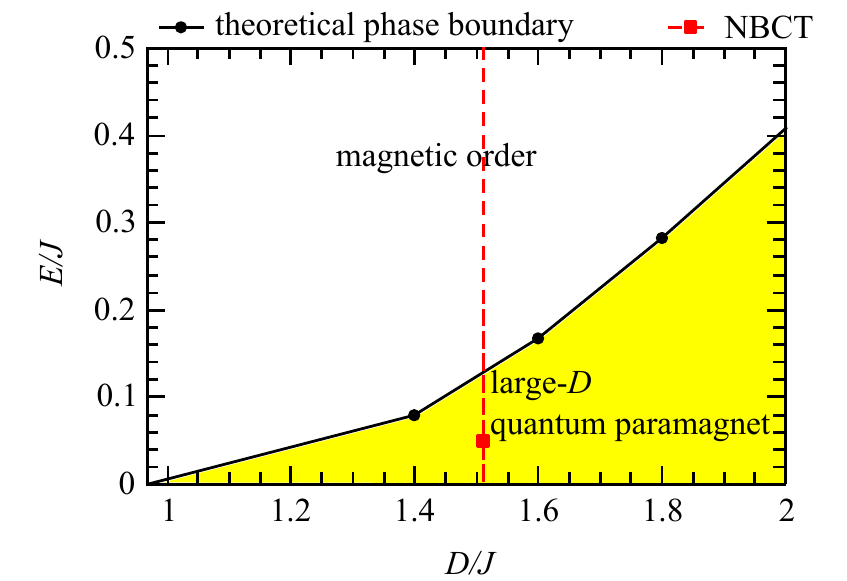}
\caption{\label{fig:phase_diagram} Anisotropic spin chain phase diagram.  The theoretical phase boundary is from reference \cite{Tzeng2017}. The $D/J$ = 1.51 line is shown along with a marker for $E/J$ = 0.05, which places NBCT in the large-$D$ QPM phase but close to a long-range ordering transition.}
\end{figure}

To understand the position of NBCT in a spin chain phase diagram and look for possible quantum phase transitions, the magnitude of the rhombic single-ion anisotropy must be estimated.  The NENP material with $D/J$ = 0.18 was reported to have $E/D$ = 0.11 \cite{Regnault1994}.  A survey of large-$D$ QPM phase chains shows a range of $E/D$ values from 0.1 to 0.3 \cite{Rudowicz2014}.  Wavefunction theory calculations of NBCT give $D$ = 13.44~K (1.16~meV) and $E$ = 1.80~K (0.155~meV) such that $E/D$ = 0.13, while wavefunction theory calculations of a molecular analogue of NBCT give $D$ = 7.12~K (0.61~meV) and $E$ = 1.49~K (0.128~meV) such that $E/D$ = 0.21.  Inputting the experimental gap values to the perturbation theory expansion of $D$ and $E$ \cite{Golinelli1992}, gives $E$ = 0.07~meV and $E/D$ = 0.13.  Taking the experimental INS $J$ = 0.35~meV for NBCT from Table I scales $E/D$ values of 0.13 and 0.21 to $E/J$ values of 0.20 and 0.31, respectively.  These calculations clearly show the non-negligible character of $E$ in NBCT and related compounds.  Also, the relative canting of the local quantization axes for the two magnetic Ni$^{2+}$ ions in the unit cell (Fig. \ref{fig:crystal_structure}) will have the effect of a rhombic single-ion anisotropy on the collective modes and has the potential to modify from the local single-ion values.  Presumably, the splitting of the $\langle S^{x/y}S^{x/y}\rangle$ mode as a function of $E$ will eventually close the gap on the lowest mode and long-range order will appear.  If at $D/J = 1.51$ the critical $E_C/J$ is estimated as $\approx 0.11$, then a linear dependence of the energy splitting yields $E/J$ = 1.3 $\Delta_E$ for NBCT and $E/J$ = 0.07.  We performed an additional DMRG calculation with $D/J$ = 1.51 and $E/J$ = 0.07, which yield (for $J$ = 0.35~meV) $\Delta _1 = 0.046$~meV, $\Delta _2 = 0.129$~meV, and $\Delta _E = 0.083$~meV.  The INS $\Delta _E = 0.054$ ($0.05 \pm 0.01$)~meV implies $E/J < 0.07$ for NBCT and then $E/J \approx 0.05$ ($E/D \approx 0.03$) assuming that $\Delta _E \propto E/J$ near $E/J = 0.07$.  Indeed, the absence of magnetic Bragg scattering at temperatures significantly less than the exchange energy scale is in opposition to a magnetically ordered ground state for NBCT, while the low temperature specific heat \cite{Manson2012} and magnetization data \cite{Xia2018} may indicate the nearby presence of strong quantum fluctuations.  Finally, NBCT may be plotted on the  ($D/J$, $E/J$) phase diagram \cite{Tzeng2017}, Fig. \ref{fig:phase_diagram}, and is close to the predicted boundary between a large-$D$ phase and a magnetically ordered phase.

\section{\label{conclusions}Conclusions and Open Questions\protect}
The magnetic excitations in isotopically enriched [Ni(HF$_2$)(3-Clpyridine)$_4$]BF$_4$ (NCBT) have been measured at $T$ = 0.3~K using inelastic neutron scattering techniques on a sample of randomly arranged microcrystals.  These data are combined with numerical density-matrix-renormalization-group studies to indicate $D/J$ = 1.51 and $E/J \approx$ 0.05, and these results place NBCT in the large-$D$ quantum paramagnetic phase but exceedingly close to a long-range ordering transition.  These findings are significantly different than the initial interpretations that this system was in the Haldane phase with $D/J$ $\approx$ 0.88 \cite{Manson2012}.  Consequently, NBCT is a prime candidate for pressure or doping induced quantum criticality.  Moreover, the importance of rhombic-anisotropy in regions close to the $D_C/J$ critical points is emphasized by NBCT.  Additionally, doping or grinding experiments to look for end-chain spins and confirm or disconfirm the proposed non-Haldane ground-state for NBCT will be useful, as would specific heat measurements down to 0.3~K or less.  The synthesis of sizable ($>$100 mg) single crystals would also allow for more investigation of NBCT at low temperature and in finite magnetic fields. 

\begin{acknowledgments}
D.~M.~Pajerowski and A.~P.~Podlesnyak are supported through the Scientific User Facilities Division of the Department of Energy (DOE) Office of Science, sponsored by the Basic Energy Science (BES) Program, DOE Office of Science.  This research used resources at the Spallation Neutron Source, a DOE Office of Science User Facility operated by the Oak Ridge National Laboratory (ORNL).  J.~Herbrych was supported by the US DOE Office of Science, sponsored by the BES Program in Materials Sciences and Engineering Division of ORNL and by the Polish National Agency of Academic Exchange (NAWA) under contract PPN/PPO/2018/1/00035.  Aspects of this work were partially supported by funding provided by the National Science Foundation via DMR-1703003 (JLM) and DMR-1708410 (MWM).  A portion of this work was performed at the National High Magnetic Field Laboratory, which is supported by National Science Foundation Cooperative Agreement No. DMR-1644779 and the State of Florida.  We acknowledge C. Batista, A. Tsvelik, and G. E. Granroth for conversations about analyzing the data.  We acknowledge M. Whangbo for a calculation of the single-ion anisotropy and G.~Alvarez for developing the DMRG++ code (https://g1257.github.io/dmrgPlusPlus/).  This manuscript has been authored by UT-Battelle, LLC under Contract No. DE-AC05-00OR22725 with the U.S. Department of Energy.  The United States Government retains and the publisher, by accepting the article for publication, acknowledges that the United States Government retains a non-exclusive, paid-up, irrevocable, world-wide license to publish or reproduce the published form of this manuscript, or allow others to do so, for United States Government purposes.  The Department of Energy will provide public access to these results of federally sponsored research in accordance with the DOE Public Access Plan (http://energy.gov/downloads/doe-public-access-plan).
\end{acknowledgments}

\appendix

\section{Technical details}

All reactions for NBCT synthesis were carried out using plasticware.  $^{11}$B(OH)$_3$ was obtained from Aldrich Chemical and 3-chloropyridine-D$_4$ (98.9\% atom \%D) was purchased from CDN Isotopes. Each was used without purification.

$H^{11}BF_4$. A 4.000 g (64.5 mmol) mass of $^{11}$B(OH)$_3$ was dissolved in 6.5 mL of HF (aq) (d = 1.16 g/mL; 258 mmol) while stirring to produce a theoretical yield of 5.676 g of aqueous H$^{11}$BF$_4$.

$Ni(^{11}BF_4)_2 \cdot yH_2O$. While stirring, the colorless H$^{11}$BF$_4$ solution was completely added to 3.828 g of neat NiCO$_3$ to produce a dark green solution from which CO$_2$ (g) evolved. The reaction mixture was allowed to stir for about 30 minutes until cessation of bubbling CO$_2$, leaving a clear green solution that was lighter in color. The solution was heated while stirring for approximately 2 hrs to remove the H$_2$O solvent at which point a moist emerald green solid of Ni($^{11}$BF$_4$)$_2$ was retrieved.

$[Ni(HF_2)(3-Clpy-D_4)_4]^{11}BF_4$. Ni($^{11}$BF$_4$)$_2 \cdot$yH$_2$O (0.7917 g, 3.40 mmol) and NH$_4$HF$_2$ (0.1939 g, 3.40 mmol) were dissolved together in 4-mL of H$_2$O to produce a light green solution. This solution was added to neat 3-chloropyridine-D$_4$ (2.000 g, 17.01 mmol) to yield a turquoise colored solution. The plastic beaker was covered with perforated parafilm and allowed to slowly evaporate at room temperature. After four days, a large mass of turquoise blue solid had formed which was collected by vacuum filtration to afford 1.4082 g of microcrystalline product. The product identity was established by infrared spectroscopy and single-crystal X-ray diffraction and shown to be the desired [Ni(HF$_2$)(3-Clpy-D$_4$)$_4$]$^{11}$BF$_4$ material.  The low-field magnetic response of a small sample (nominally 15 mg) was studied using a commercial SQUID magnetometer (MPMS-XL) and, to within experimental uncertainties, the measured temperature dependence (5~K $\leq$ T $\leq$ 300~K) was the same as reported Ref. \cite{Manson2012}.

For the neutron experiments, $\approx$1 gram of powder was mounted in an aluminum can with a copper lid.  Cryogenic temperatures were achieved with a wet $^3$He cryostat for the $T$ = 0.3~K and $T$ = 13~K data, and a cryogen-free dilution fridge for the $T$ = 0.07~K data.  The time-of-flight spectrometer at the SNS BL-5 cold chopper neutron spectrometer (CNCS) was used in high-flux mode \cite{CNCS1}.  Experimental energy resolutions are from MANTID \cite{Mantid}, and at $\hbar\omega$ = 0 the full-width-half-max resolution 0.02~meV and 0.11~meV for 1~meV and 3.32~meV incident energies, respectively.  The data were normalized to the proton current on target during collection.  The detectors were normalized using a vanadium standard measurement that is defined to have an average intensity per pixel of 1 scaled unit.  Intensities are multiplied by the ratio of incident and final momentum to result in numbers that are proportional to a correlation function.  All numerical optimizations used the libraries of SciPy \cite{Jones2001}.

The DMRG calculations were performed for $D/J$ = [0, 0.5, 1.0, 1.5, 2.0] using L = 64 sites on a chain.  For $\omega/J < 0.2$, $\Delta\omega/J = 0.002$ and $\eta = 0.0032$.  For $\omega/J > 0.2$, $\Delta\omega/J = 0.04$ and $\eta = 0.07$.  For de-noising, these DMRG correlations were fit to the phenomenological relationships
\begin{equation}
\begin{aligned}
    I(q_{1D},\hbar \omega) = (1 - cos(q_{1D})) (I_0 + \frac{I_{-1}}{\hbar \omega}),\\
    \frac{E(q_{1D})}{J} = \sqrt{A cos(\frac{q_{1D}}{2})^2 + v^2 sin^2(q_{1D}) + \Delta^2},
\label{eq2}
\end{aligned}
\end{equation}
where the dispersion relationship parameterization is inspired by linear spin wave theory \cite{Ma1992}.  The resulting parameters were then interpolated with cubic splines to give smooth functions of $A$, $v$, $\Delta$, $I_0$, and $I_{-1}$ for the $\langle S^{x/y} S^{x/y} \rangle$ and $\langle S^zS^z \rangle$ correlation functions.  The $\langle S^{x/y} S^{x/y} \rangle$ correlations were further split into $\langle S^{x} S^{x} \rangle$ and $\langle S^{y} S^{y} \rangle$ by letting $\Delta_{x/y} \rightarrow \Delta_{x/y} \pm \Delta_E$/2.  To compare DMRG with experiment, a Bose factor was included to modify the intensity as a function of $\hbar\omega$.  The Ni$^{2+}$ magnetic form factor was included to modify the intensity as a function of momentum transfer \cite{Borner2003}.  Subsequent to fitting with the $D/J$ = [0, 0.5, 1.0, 1.5, 2.0] interpolated DMRG data, a $D/J$ = 1.51 DMRG calculation was performed that agreed with the interpolated data to within 1 part per 1,000 and therefore has no effect on the solution.

The NBCT and the mononuclear analog [Ni(3-Cl-py)$_4$(FHF)$_2$] were modeled by wavefunction theory using a complete active space self-consistent-field (CASSCF) approach employing the scalar relativistically recontracted basis sets tailored for use with the Douglas-Kroll-Hess (DKH) Hamiltonian. Calculations were done in the Orca program suite. \cite{Neese2012} The basis sets were of triple-$\zeta$ quality on all atoms (DKH- def2-TZVP) except for Zr where the “old-DKH-TZVP” of Orca was employed. In all calculations, the second order DKH Hamiltonian was used. The calculation was done on the experimental geometry and the active space was limited to the Ni d-orbitals, leading to a CAS(8,5) calculation.

\begin{figure*}
\includegraphics{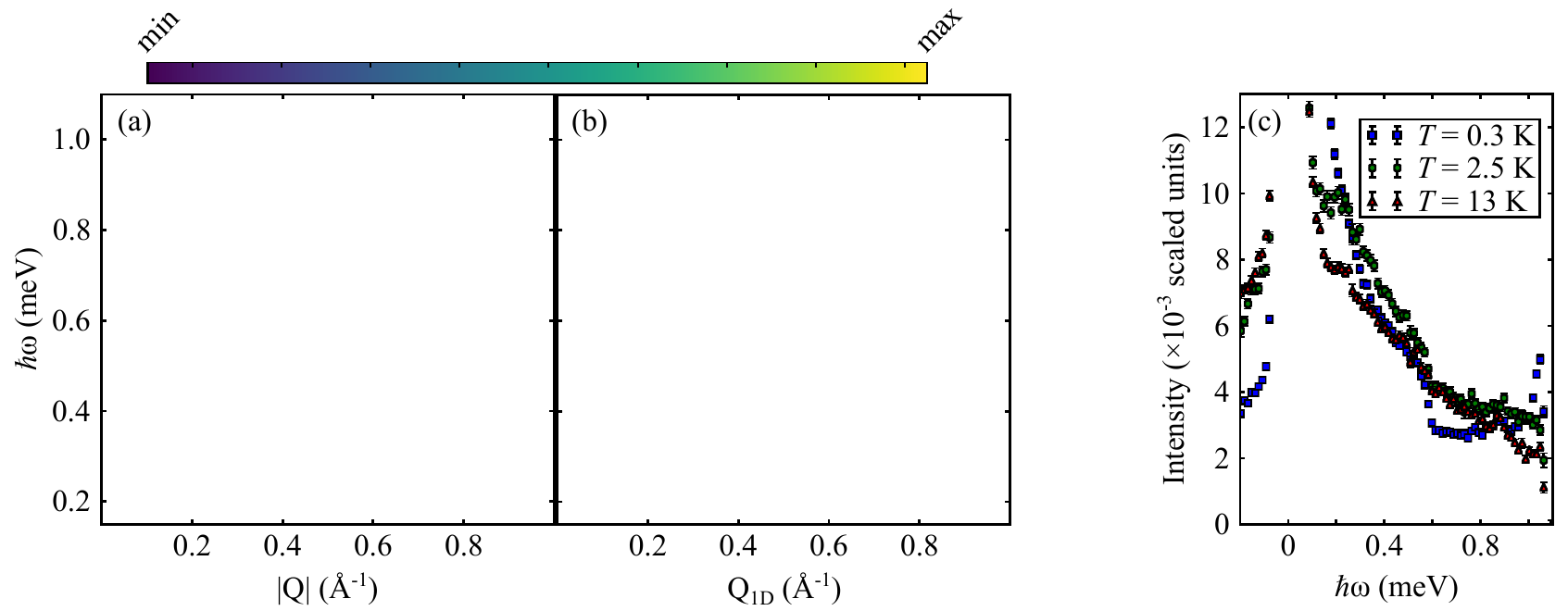}
\caption{\label{fig:powder_conversion_and_local_mode}Local mode in NBCT.  (a) E$_i$ = 1.55~meV, $T$ = 0.3~K data, with min = $0$ and max = $2\times10^{-2}$ intensity scale.  (b) E$_i$ = 1.55~meV, $T$ = 0.3~K data minus $T$ = 13~K data with = $–1\times10^{-5}$ and max = $1\times10^{-4}$ intensity scale.  In (a) and (b), white regions are where the scattering condition is not satisfied by the spectrometer.  (c) Intensity versus energy transfer for E$_i$ = 1.55~meV powder averaged data near $q_{1D} = \pi$ ($Q_{1D}$ averaged between [0.4, 0.6] Å$^{-1}$).}
\end{figure*}
\section{Local mode and example of I($Q_{1D}$) extraction}

The local mode we assign to single-ion $D$ excitations from Ni$^{2+}$ spins without spatial correlation is visible in the raw powder data at $T$ = 0.3~K, Fig. \ref{fig:powder_conversion_and_local_mode} (a).  At low momentum transfers, contaminating intensity from the direct beam is visible.  Multiple scattering between the sample and the cryostat is also visible, such as at $(|Q|, \hbar \omega) \approx (1~$Å$^{-1}, 0.6$~meV$)$.  Here, we also visualize the extraction of the one-dimensional correlations from the powder averaged data.  Subtracting $T$ = 13~K data from $T$ = 0.3~K data removes extrinsic features, and subsequently converting from $|Q|$ to $Q_{1D}$ removes the bleeding of intensity to higher momenta, Fig. \ref{fig:powder_conversion_and_local_mode} (b).  The temperature dependence of the raw powder data averaged over $|Q|$ = [0.4, 0.6] Å$^{-1}$ shows the evolution of both the one-dimensional collective modes and the local mode, Fig. \ref{fig:powder_conversion_and_local_mode} (c). 


%

\end{document}